\documentclass[twocolumn,floatfix,amsmath,aps,pre,showpacs]{revtex4}

\usepackage{epsfig}
\usepackage{amsmath}
\usepackage{graphicx}
\usepackage{amsmath}
\usepackage{amsthm}
\usepackage{multirow}
\usepackage{relsize}
\usepackage{amssymb}
\usepackage{bm}
\usepackage{textcomp}

\usepackage{epstopdf}
\usepackage{soul}
\RequirePackage{color}
\definecolor{MyDarkGreen}{rgb}{0.02,0.60,0.06}

\DeclareMathAlphabet{\mathitbf}{T1}{cmr}{bx}{it}

\begin{document}

\title{Computation of the dynamic critical
  exponent of the three dimensional Heisenberg model}

\author{A. Astillero} \affiliation{Departamento de Tecnolog\'ia
  de los Computadores y las Comunicaciones, Universidad de
  Extremadura, 06800 M\'erida, Spain}

\affiliation{Instituto de Computaci\'{o}n Cient\'{\i}fica Avanzada
  (ICCAEx), 06071 Badajoz, Spain.}

\author{J.J. Ruiz-Lorenzo} \affiliation{ Departamento de F\'{\i}sica,
  Universidad de Extremadura, 06071 Badajoz, Spain.}

\affiliation{Instituto de Computaci\'{o}n Cient\'{\i}fica Avanzada (ICCAEx),
  Universidad de Extremadura, 06071 Badajoz, Spain.}

\affiliation{ Instituto de Biocomputaci\'{o}n y F\'{\i}sica de
  Sistemas Complejos (BIFI), 50018 Zaragoza, Spain.}

\date{\today}

\begin{abstract}
  
Working in and out of equilibrium and using state-of-the-art
techniques we have computed the dynamic critical exponent of the three
dimensional Heisenberg model. By computing the integrated autocorrelation
time at equilibrium, for lattice sizes $L\le 64$, we have obtained
$z=2.033(5)$.  In the out of equilibrium regime we have run very large
lattices ($L\le 250$)  obtaining $z=2.04(2)$ from
the growth of the correlation length.  We compare our values with that
previously computed at equilibrium with relatively small lattices
($L\le 24$), with that provided by means a three-loops calculation
using perturbation theory and with experiments.  Finally we have
checked previous estimates of the static critical exponents, $\eta$
and $\nu$, in the out of equilibrium regime.

\end{abstract}


\maketitle

\section{Introduction}\label{I}

The study of the dynamics in and out of equilibrium in a critical
phase is of paramount importance since it permits to extract the
critical exponents of the system, hence, to characterize its
universality class. In the last decades a great amount of work,
analytical, numerical and experimental, has been devoted to study
these issues.

One of the studied systems has been the three dimensional (isotropic)
classical Heisenberg model.
The dynamic critical exponent, $z$, has been computed using field
theory by studying its Model A dynamics (pure relaxational dynamics in
which the order parameter is not
conserved).\cite{Hohen:77,Folk:06,Tauber:17} A three-loop computation
reported in Ref.~\cite{Antonov:84} provided $z=2.02$.~\footnote{ $ z =
  2 + c \eta \,\,,\,\, c = 0.726-0.137 \epsilon + O(\epsilon^2)\,, $
  where $\eta$ is the anomalous dimension of the field (from static)
  and $\epsilon=4-d$, $d$ being the dimensionality of the model.}
The equilibrium dynamics of this model was studied by means of
numerical simulations in Ref.~\cite{Peczak:93} and $z=1.96(6)$ was
reported. The authors were aware that this exponent was slightly below
the analytical computation of Ref.~\cite{Antonov:84} and discuss in
the paper different systematic bias. For example, a relatively narrow
range of the lattice sizes and despite the accuracy of their values
for the correlation times, a more precise determination of these times
were needed to study the corrections to the scaling presented in the model.

From the experimental side, the situation is complicated due to
the crossover from the Heisenberg universality class to the dipolar
one which induces a change from $z\sim 2.5$ (Heisenberg with
conserved magnetization and reversible forces, model
J~\cite{Hohen:77,Folk:06,Tauber:17}) to $z\sim 2$
(dipolar).~\cite{Chow:80,Hohenemser:83}  In particular using
PAC~\footnote{Perturbed Angular Correlations of $\gamma$ ray
  spectroscopy.}  Hohenemser {\em et al.} found
$z=2.06(4)$~\cite{Hohenemser:83} for Ni and $z\simeq 2$ for Fe; 
Dunlap {\em et al.}~\cite{Dunlap:80} using ESR~\footnote{Electron Spin
  Resonance.} found $z=2.04(7)$ for EuO; and $z=2.09(6)$ was found by Bohn {\em
  et al} for EuS~\cite{Bohn:84} using inelastic neutron scattering. It
seems that the interplay of spin dipoles with orbital angular momentum
or dipolar interactions breaks the conservation of the magnetization on
these materials, producing a crossover between Heisenberg model J
($z\sim 2.5$) and Heisenberg model A ($z\sim
2$).~\cite{Chow:80,Hohenemser:83}

Recently, Pelissetto and Vicari~\cite{Peli:16} have used the value
provided by field theory in the scaling analysis of their numerical
data to study the off-equilibrium behavior of three-dimensional $O(N)$
models driven by time-dependent external fields and assigned it an
error of 0.01, so $z=2.02(1)$, to take into account the uncertainty on
the extrapolation to $\epsilon=1$ of the three-loop-expansion result.

Consequently, it is of paramount importance to obtain an accurate value for
this dynamic critical exponent, in order to be used in future
numerical analysis and experiments, and also to check the accuracy of the
three-loops analytical computation. The main goal of our study is to
improve the value of $z$ using numerical simulations by studying the
dependence of the integrated correlation time in the equilibrium
regime and the behavior of the correlation length, susceptibility and
energy with the simulation time in the out of equilibrium 
region. In this way, we can compare the performance of equilibrium and
out-of-equilibrium methods in the computation of the dynamic critical
exponent.

Nowadays, a great amount of work has been devoted to study numerically
the dynamics of disordered systems, see for example Refs.
\cite{Parisi:99,Hasen:07,Janus1,Janus2,Lulli:16,Janus3,Janus4}. In
general a sudden quenched is performed to work in the off-equilibrium,
yet, in other studies the models have been simulated at
equilibrium. For example in the three dimensional diluted Ising model
both approaches gave the same dynamic critical exponent.~\cite{Hasen:07}

In the equilibrium part of the paper we compute the integrated
correlation time, avoiding some of the problems which appear in the
computation of the exponential one (e.g. assume that the
autocorrelation function is a single exponential
function~\cite{Sokal,Salas}).

We also study the correlation length in the out-of-equilibrium
regime. In the last two decades, this observable has played an
important role both in numerical simulations~\cite{Janus1,Janus2} and
experiments out of equilibrium~\cite{Orbach:14} in spin glasses. Due
to this, powerful numerical techniques has been developed to compute
this observable with high accuracy which has allowed a precise
determination of the dynamic critical exponent just at the critical
point as well as inside the critical spin glass
phase.~\cite{Janus3} We apply these techniques to
the three dimensional (non disordered) Heisenberg model.  In addition
to the computation of the dynamic critical exponent, we have checked
the consistency of previous and very accurate determinations of the
static critical exponents ($\nu$ and $\eta$) in the out-of-equilibrium
regime.  Our starting point will be the (very precise) critical
temperature computed in Ref. \cite{Balles:96} and the static critical
exponents reported in Refs.~\cite{Campos:02,Hasen:11}.

We have also measured the dynamic critical exponent from the
decay of the energy at criticality. This decay has also been studied in
the past in finite dimensional spin glass~\cite{Janus1} and recently
has played a central role together with the behavior of the
correlation length in the analysis of the Mpemba effect, a striking
memory effect.~\cite{Janus4}

The structure of the paper is the following. In the next section we
introduce the model and the observables. In section \ref{sec:num}
we describe our numerical results: in Sec. \ref{sec:equi} we 
report the equilibrium determination of $z$ via the integrated
autocorrelation time; in Sec. \ref{sec:zeta} we study the
dependence of the correlation length with time and the computation of
$z$ out of equilibrium; in Sec. \ref{sec:eta} and Sec. {\ref{sec:ener}
  the correlation function and the energy have been studied
  (respectively). Sec. \ref{sec:concl} is devoted to the
  conclusions. Two appendices  close the paper, one to describe
  our implementation of GPU and the last one to describe how we have
  computed the statistical error of the exponents with highly
  correlated data.

\section{The model and Observables}\label{MO}
The Hamiltonian of the three dimensional Heisenberg model is
\begin{equation}
  {\mathcal H} =  - \sum_{<{\mathitbf r},{\mathitbf r'}>}
  \mathitbf{S}_{\mathitbf r}\cdot \mathitbf{S}_{\mathitbf r'} \, .
\label{origham}
\end{equation}
$\mathitbf{S}_{\mathitbf r}$ is a classical three component spin on
the site $\mathitbf{r}$ of a three dimensional cubic lattice with
volume $V=L^3$ and periodic boundary conditions. Without loss of
generality we will assume that the spins are unit vectors. The sum
runs over all pairs of nearest neighbors spins. We have simulated this
model with the standard Metropolis algorithm~\footnote{In the
  equilibrium simulations, Sec. \ref{sec:equi}, we have used the
  Metropolis algorithm proposing a random spin in the unit sphere. In
  the out of equilibrium simulations we have used the Metropolis
  algorithm in the standard way~\cite{AmitMartinMayor}: we modify the
  original spin by adding a random vector and normalizing the final
  vector to the unit sphere. The magnitude of the random vector is
  selected in order to maintain an acceptance between 40\% and
  60\%. Both versions of the Metropolis algorithm belong to the same
  dynamic universality class.\cite{Sokal}} and we have run in CPU
(smaller time simulations) and GPU (for larger time
simulations). Details on the simulations can be found in the appendix
A.

\subsection{Equilibrium}
\label{sec:defequil}

We address the problem of the computation of the dynamic critical
exponent in the equilibrium regime by means the computation and
further analysis of the integrated aucorrelation time as a function of
the lattice size.

We compute for a given observable ${\cal O}(t)$,  the autocorrelation
function (we follow Refs. \cite{Madras,Sokal,AmitMartinMayor}):
\begin{equation}
  \label{eq:auto}
  C_{\cal O}(t)\equiv \langle {\cal O}(s) {\cal O}(t+s) \rangle -\langle {\cal O}(t) \rangle^2\,,
\end{equation}
and its normalized version
\begin{equation}
  \label{eq:autoNorm}
  \rho_{\cal O}(t)\equiv C_{\cal O}(t)/C_{\cal O}(0)\,.
\end{equation}
The integrated autocorrelation time is given by
\begin{equation}
  \label{eq:tau}
\tau_{\mathrm{int},{\cal O}}= \frac{1}{2}+ \sum_{t=0}^\infty \rho_{\cal O}(t)\,.
\end{equation}
In a run with $N$ measurements, the number of independent measurements
of the observable ${\cal O}$ is just $N/(2 \tau_{\mathrm{int},{\cal
    O}})$.~\cite{Madras,Sokal,AmitMartinMayor} If the number of
measurements is finite, for large times $t$ the noise will dominate
the signal in $\rho_{\cal O}(t)$ and to bypass this problem
we  use the following
self-consistent method to compute the integrated time
\begin{equation}
  \label{eq:tauM}
\tau_{\mathrm{int},{\cal O}}= \frac{1}{2}+ \sum_{t=0}^{ c \tau_{\mathrm{int},{\cal O}}} \rho_{\cal O}(t)\,,
\end{equation}
where $c$ is usually
taken to be 6 or bigger.~\cite{Madras,Sokal,AmitMartinMayor}

At the critical point the integrated autocorrelation time of a long
distance observable diverges with the size of the
system~\cite{Hasen:07}
\begin{equation}
  \label{eq:tauscaling}
\tau_{\mathrm{int},{\cal O}}\sim L^z \large(1+  O(L^{-\omega})\large)\,.
\end{equation}
where $z$ is the dynamic critical exponent and $\omega$
is the leading correction-to-scaling exponent (the leading irrelevant
eigenvalue of the theory).

Another time, the exponential correlation time, is  defined as
\begin{equation}
\tau_\mathrm{exp,{\cal O}} \equiv \limsup_{t\to \infty} \frac{-t}{\log \rho_{\cal O}(t)} \,,
\end{equation}
which also depends on the observable used to define $\rho_{\cal O}(t)$. The
exponential autocorrelation time controls the approach to the
equilibrium.\cite{Sokal}

Once we have defined the exponential correlation function we can
write the general scaling form for the correlation function~\cite{Sokal}
\begin{equation}
  \label{eq:CorScaling}
\rho_{\cal O}(t)= t^{-p_{\cal{O}}} f_{\cal{O}}\Big(\frac{t}{\tau_{\mathrm{exp},{\cal O}}}, \frac{\xi(L)}{L}\Big)\,,
\end{equation}
where $\xi(L)$ is the equilibrium correlation length computed on a
system of size $L$.  Integrating Eq.~(\ref{eq:CorScaling}) in time, we
obtain the integrated correlation time and that $\tau_\mathrm{int}\sim
\tau_\mathrm{exp} ^{1-p_{\cal{O}}}$. Both times are proportional if
and only if $p_{\cal{O}}=0$ and in this situation $z$ is the same for
both times. Otherwise, $p_{\cal{O}}\neq 0$, and $\tau_\mathrm{exp}$
and $\tau_\mathrm{int}$ will provide with different dynamic critical
exponents. See also the discussion of Ref. \cite{Hasen:07}.

In this paper we will use the slowest mode provided by the non local
operator ${\cal O}=\mathitbf{M}^2$, where the magnetization
${\mathitbf M}$ is defined as
\begin{equation}
{\mathitbf M} = \sum_{\mathitbf x} {\mathitbf S}_{\mathitbf x}\,.
\end {equation}

\subsection{Out of equilibrium}

We have focused on only one local observable, the energy, defined as
\begin{equation}
e(t)=\frac{\langle {\mathcal H} \rangle_t}{V}\,.
\end{equation}
We denote the average over different initial conditions at the Monte
Carlo time $t$ by $\langle(\cdots)\rangle_t$. The renormalization
group predicts~\cite{AmitMartinMayor,Tauber:17}, at the critical point,
the following behavior for this observable:
\begin{equation}
e(t)=e_\infty+ C t^{(d-1/\nu)/z} \left(1+ A t^{-\omega/z} \right)\,,
\end{equation}
where $d$ is the dimensionality of the space (three in this study) and
$\nu$ is the critical exponent which controls the divergence of the
equilibrium correlation length. The other two exponents $z$ and
$\omega$ has been defined in the previous subsection.

One of the main observables on this paper is the correlation function defined as:
\begin{equation}
  \label{eq:cor}
  C(r,t)=\frac{1}{V} \sum_{\mathitbf x}
  \langle {\mathitbf S}_{\mathitbf x} {\mathitbf S}_{\mathitbf{r+x}} \rangle_t\,,
\end{equation}
satisfying, at criticality, the 
following scaling law~\cite{Tauber:17}
  \begin{equation}
C(r,t)=\frac{1}{r^a} f\left(\frac{r}{\xi(t)}\right)
  \end{equation}
which defines the dynamic correlation length, $\xi(t)$. As we
approach the equilibrium regime, $\xi(t)$ reaches its equilibrium
value.

At the $d=3$  critical point 
and in equilibrium, one should expect
  \begin{equation}
C(r,t)\sim\frac{1}{r^{d-2+\eta}} = \frac{1}{r^{1+\eta}} \,,
\label{eq:corre-equi}
  \end{equation}
$\eta$ being the anomalous dimension of the field.
  
The correlation length $\xi(t)$ can be estimated by computing~\cite{Janus1,Janus2}
\begin{equation}
  I_k(t)=\int_{0}^{L/2} dr~ r^k C(r,t)\,,
\label{eq:I}
\end{equation}
by means of
\begin{equation}
  \xi_{k,k+1}(t)\equiv \frac{I_{k+1}(t)}{I_k(t)}\,.
\end{equation}
We focus in this work on $\xi_{2,3}$. On spin glasses
was measured $\xi_{1,2}$ with a correlation function
decaying like $1/r^{0.5}$.~\cite{Janus1,Janus2} In our case, to decrease the
weight of the smallest distances we have resorted to compute higher values
of $I_k$. In the appendix B we describe the detailed procedure we
have used to compute the integrals and how we have estimate the
statistical error associated with $\xi_{k,k+1}(t)$. The dependence of
the dynamic correlation length with time is
\begin{equation}
  \xi_{k,k+1}(t)\sim t^{1/z} \left(1+ A_k t^{-\omega/z}  \right)\,.
  \label{eq:xiNL}
\end{equation}
The magnetic susceptibility is given by
\begin{equation}
  \chi(t)=\frac{1}{V} \langle {\mathitbf M}^2 \rangle_t\,,
\end{equation} 
or equivalently by
\begin{equation}
  \chi(t)=\int d^3 x~ C(|\mathitbf{x}|,t) \,.
\end{equation}
In the regime of large $\xi(t)$ we recover rotational invariance and we obtain
\begin{equation}
  \chi(t)=4 \pi I_2(t)\,.
\end{equation}
The temporal dependence of $\chi(t)$ is
\begin{equation}
  \chi(t)\sim t^{(2-\eta)/z} \left(1+ A t^{-\omega/z}  \right)\,\,,
\end{equation}
which can be rewritten as
\begin{equation}
  \chi(t)\sim \xi_{k,k+1}(t)^{2-\eta} \left(1+ C_k \xi(t)^{-\omega}  \right)\,\,.
\end{equation}

\section{Numerical results}
\label{sec:num}
In this section we report the computation of the integrated correlation
time at equilibrium.
After this analysis, we describe our results in the out of
equilibrium regime. In particular, we consider the short and long
time behavior of correlation length and the long time behavior of the
correlation function and that of the energy.  The data are obtained
after a sudden quench from $T=\infty$ to $T=1/\beta_c$.
All the numerical simulations were performed at
$\beta_c=0.693001$.~\cite{Balles:96}

\subsection{Equilibrium}
\label{sec:equi}

To obtain the dynamic critical exponent in the equilibrium regime, we
compute the integrated correlation time of $\mathitbf{M}^2$ when the
numerical simulation has reached the equilibrium. We follow the
methodology described in Sec. \ref{sec:defequil}, using the self
consistent windowm algorithm with a window size given by $c
\tau_\mathrm{int}$. We have analyzed the correlation functions with
$c=6$, 8, 10 and 12 and we have checked that the $c=10$ data are fully
compatible with that of $c=8$ and 12. We report in the following
$c=10$ integrated autocorrelation times.

In Table \ref{table:tau} we report the values of
$\tau_{\mathrm{int},\mathitbf{M}^2}$ and 
other parameters of the performed runs.  In order to improve the
statistics on $\tau_{\mathrm{int},\mathitbf{M}^2}$ we have performed
50 independent runs (initial conditions). We have computed the
statistical error on the integrated autocorrelation times by using the
jackknife method over the independent runs.~\cite{jackknife,Young}

\begin{table}
\begin{ruledtabular}
\begin{tabular}{ccc}
$L$ & $\tau_{\mathrm{int},\mathitbf{M}^2}$ & $n_\mathrm{sweep}/\tau_{\mathrm{int},\mathitbf{M}^2}$  \\
  \hline
8  & 24.84(1) & 4122383 \\
12 & 53.11(4) & 1928074 \\
16 & 93.57(8) & 1094368 \\
24 & 211.2(3) &  484848 \\
32 & 378.3(5) &  270542 \\
48 & 860(3)   &   58333 \\
64 & 1545(9)  &   13782 \\
\end{tabular}
\end{ruledtabular}
\caption{Integrated correlation time of $\mathitbf{M}^2$,
  $\tau_{\mathrm{int},\mathitbf{M}^2}$ for $c=10$, as a function of the lattice
  size, $L$. We also report the length of the run at equilibrium,
  $n_\mathrm{sweep}$, in units of
  $\tau_{\mathrm{int},\mathitbf{M}^2}$. For each lattice size we
  have performed 50 intial conditions. Notice that all the reported
  runs satisfy
  $n_\mathrm{sweep}>10000~\tau_{\mathrm{int},\mathitbf{M}^2}$.\cite{Sokal}
  \label{table:tau}}
\end{table}  

We have fitted $\tau_{\mathrm{int},\mathitbf{M}^2}$ to
Eq. (\ref{eq:tauscaling}) using $8\le L \le 64$ obtaining $z=2.033(5)$
and $\omega=2.7(3)$ with a $\chi^2/\mathrm{d.o.f.}=0.36/3$.\footnote{In
  this case, the data for different lattice sizes are not correlated
  and we can safely use the diagonal covariance matrix.} We report
this fit and the numerical data in Fig. \ref{fig:equil}.  Fitting the
data using only a power law (i.e. neglecting the correction-to-scaling term) we
obtain a good fit only for $L\ge 24$ obtaining $z=2.026(4)$ with
$\chi^2/\mathrm{d.o.f.}=0.28/2$.  Both reported values are fully compatible.

\begin{figure}[h!]
\centering
\includegraphics[width=\columnwidth, angle=0]{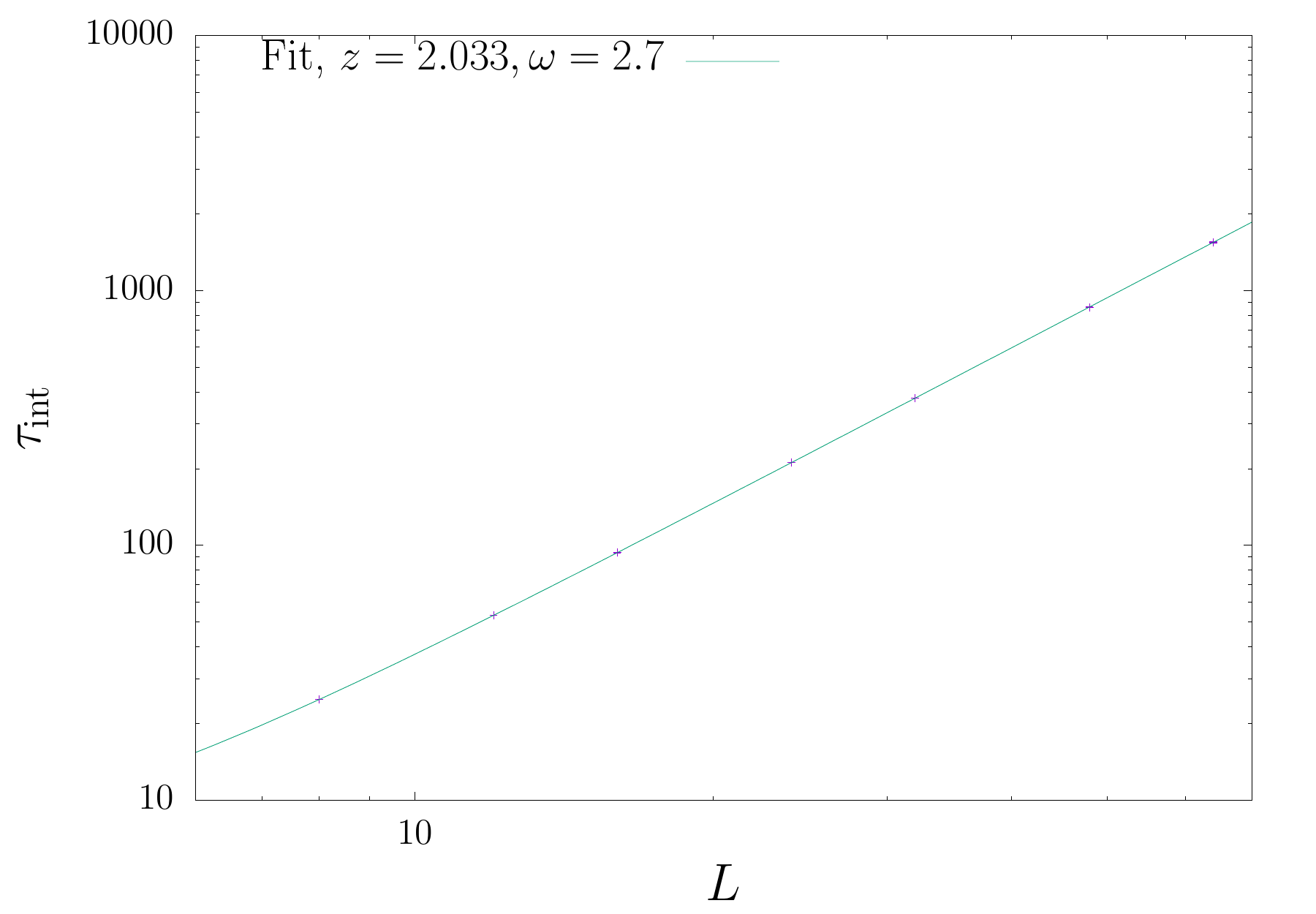}
\caption{(color online) Behavior of the integrated correlation time,
  $\tau_{\mathrm{int},\mathitbf{M}^2}$, as a function of the lattice size, $L$. We
  have also shown our best fit taking into account corrections to the 
  scaling (see the text).}
\label{fig:equil}
\end{figure}

Finally we present a scaling analysis of the $\rho(t)$ function at
equilibrium and at the critical point to show that $\tau_\mathrm{exp}$
and $\tau_\mathrm{int}$ are proportional and therefore both times
diverge with the same dynamic critical exponent
$z$. Fig. \ref{fig:CorScaling} shows the scaling law of the
correlation function $\rho(t)$ as a function of
$t/\tau_\mathrm{int}(L)$ instead of $t/\tau_\mathrm{exp}(L)$, as
stated in Eq. (\ref{eq:CorScaling}). Scaling in the new variable holds
if and only if $\tau_\mathrm{exp} \propto \tau_\mathrm{int}$ and this
is the case apart from small scaling corrections on the $L\le 16$ data
induced by the term $\xi(L)/L$ in the scaling function $f_{\cal{O}}$
of Eq. (\ref{eq:CorScaling}), see also the inset of this figure for a
detailed and more quantitative view of this effect.

\begin{figure}[h!]
\centering
\includegraphics[width=\columnwidth, angle=0]{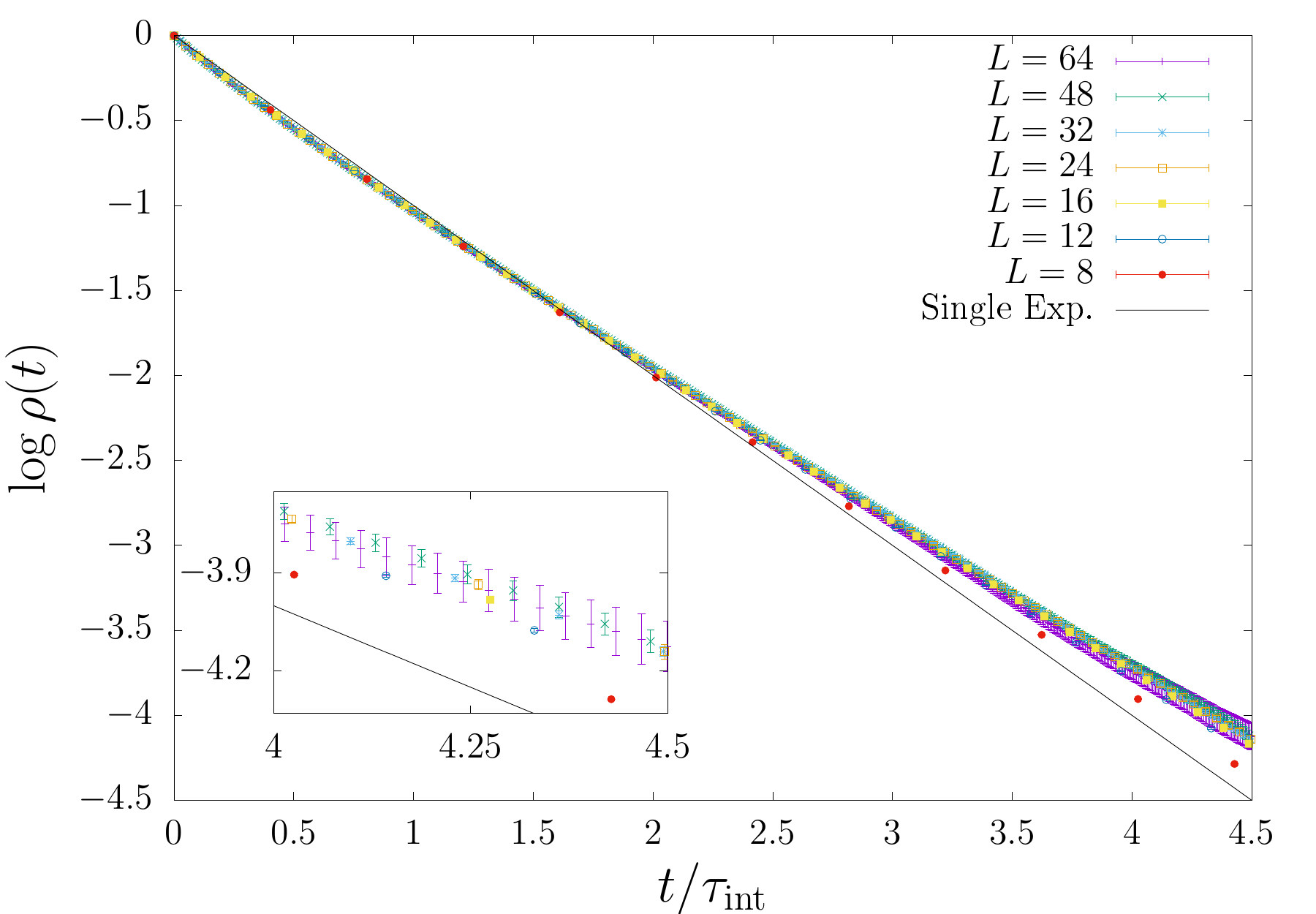}
\caption{(color online) Behavior of the integrated correlation
  function, $\rho(t)$ for the squared magnetization as a function of
  $t/\tau_\mathrm{int}(L)$.  Inset: we show a zoom of the long time region of 
  the main plot, drawing only a small number of points, so the reader
  can see in a better way the differences among the different lattice
  sizes.  Notice that all the data with $L\ge 16$ collapse in the
  scaling formula and this fact  provides a numerical verification of the
  proportionality of the integrated and exponential correlation
  times. We have also plotted a single pure exponential, $\exp(-x)$,
  to show that the correlation function is not a single exponential.
}
\label{fig:CorScaling}
\end{figure}

In Ref. \cite{Peczak:93} the (biggest) exponential correlation time
was computed for the magnetization $\sqrt{\mathitbf{M}^2}$.  However,
the computation of this exponential time is very involved in the case
the autocorrelation function $\rho(t)$ does not show a single exponential
decay ~\cite{Salas,Peli:16}: in Fig. \ref{fig:CorScaling} we have
plotted a single exponential decay and the correlation function
clearly departs from this behavior.

\subsection{Correlation length: Shorter times}
\label{sec:zeta}

We report in Fig. \ref{fig:XI12EQUIL} the behavior of $\xi_{23}$ as a
function of time for different lattice sizes that we have  been able to
thermalize.  The $\xi_{23}$-plateaus obtained for the largest times
are a clear evidence that the numerical simulations have reached the
equilibrium.

In order to extract the dynamic critical exponent by using
Eq. (\ref{eq:xiNL}), we need to work in the out of equilibrium regime,
avoiding the transient regime and  the equilibrium domain. Therefore, we
need to check the following points:
\begin {itemize}

\item We need to avoid the transient regime between the power law
  behavior (pure out of equilibrium regime) and the plateau one
  (equilibrium one).

\item Eq. (\ref{eq:xiNL}) holds after the initial transient of the
  dynamics. Therefore, we need to fix a minimum time $t_\mathrm{min}$.

\item Finally, in order to avoid finite size effects, we need to
  compare the correlation lengths for different lattice sizes. 
\end{itemize}

\begin{figure}[h!]
\centering
\includegraphics[width=\columnwidth, angle=0]{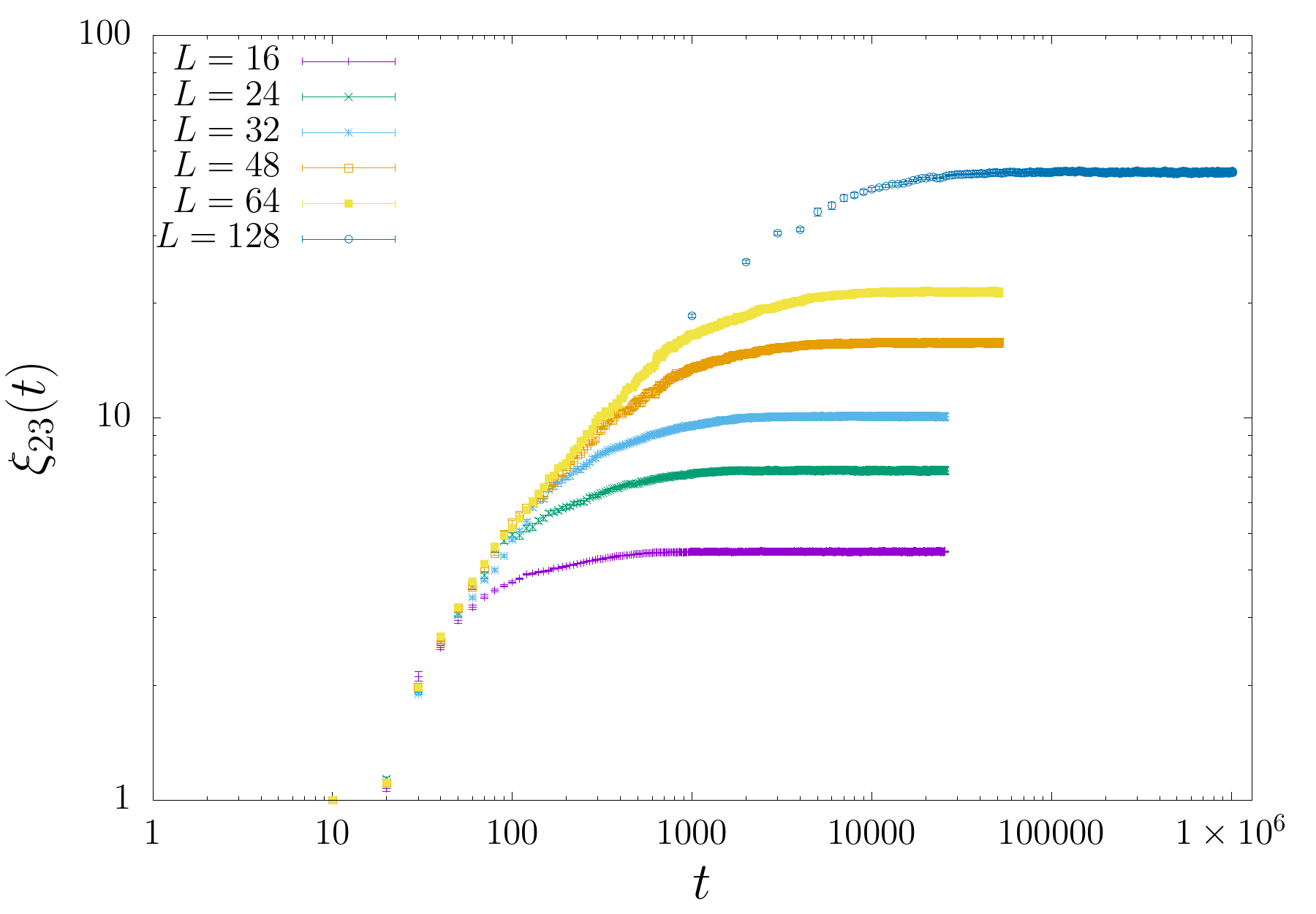}
\caption{(color online) Behavior of the dynamic correlation length for
  $L=16$, 24, 32, 48, 64 and $L=128$. We have only plotted simulations
  in which the equilibrium regime has been reached: notice the clear
  plateau of the different correlation length curves.}
\label{fig:XI12EQUIL}
\end{figure}

\begin{figure}[h!]
\centering
\includegraphics[width=\columnwidth, angle=0]{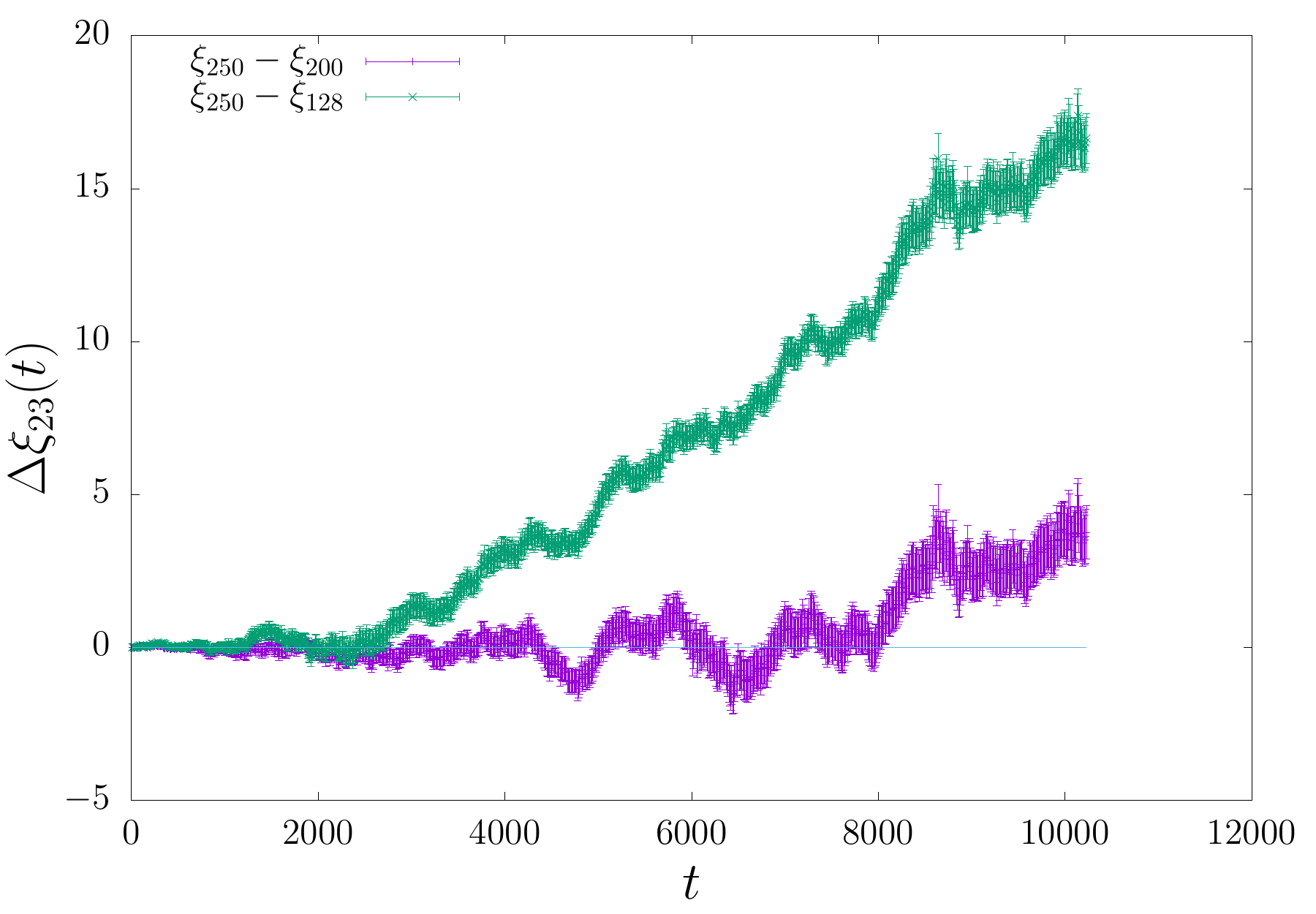}
\caption{(color online) We show the difference of dynamic correlation
  lengths, $\Delta \xi_{23}(t)$ for three pairs of lattice sizes as a
  function of time: $\xi_{L=250}-\xi_{L=200}$ and
  $\xi_{L=250}-\xi_{L=128}$.  The zero value has been marked with a
  horizontal line. Notice that the $L=250$ data are asymptotic (as
  compared with those of $L=200$) for $t<8100$ (the data lie, at most, at one
  standard deviation of the zero value).}
\label{fig:XI12DIF}
\end{figure}

In this section we have performed numerical simulations with $t<10240$
(in order to avoid the transient and equilibrium regimes). We have
simulated 4000 random initial conditions for $L=128$ and $L=200$, and
5325 initial conditions for $L=250$.

In Fig.~\ref{fig:XI12DIF} we have plotted, to check finite size
effects, the differences among the correlation lengths of $L=128$ and
$200$ and that of $250$. In this figure we can see that the data of
the $L=200$ and $L=250$ lattices are compatible in the statistical
error for $t<8100$.

From the previous discussion we must fit the data for $\xi_{23}$ in
the time interval given by $t\in [t_\mathrm{min},8100]$ using the
$L=250$ data:  $t_\mathrm{min}$ being the smallest value of the Monte
Carlo time that provides a good $\chi^2/\mathrm{d.o.f.}$ (e.g. $\sim
1$) by fitting the data to Eq. (\ref{eq:xiNL}).

In Fig. \ref{fig:XI12PURE} we show the behavior of $\xi_{23}$ for the
largest lattice size simulated in this time regime, $L=250$. By
fitting $L=250$ data in the interval $t \in [100,8100)$ we have
  obtained $z=2.04(2)$ and $\omega=2.2(4)$ with
  $\chi^2/\mathrm{d.o.f.}=287/796$.  Furthermore, we can report that a
  fit neglecting the contribution of the correction-to-scaling
  term provides $z=2.012(13)$ with $t \in [400,8100)$ and
    $\chi^2/\mathrm{d.o.f.}=361/768$. All two reported values are
    statistically compatible.

We have computed the statistical error on the $z$-exponent by means of the
jackknife method.~\cite{jackknife,Young} As described in the appendix
\ref{ApenB}, we compute the $\chi^2$ using a diagonal covariance
matrix (neglecting the correlations of the data), but we use a
jackknife procedure to take into account the (important) different
correlations among the data. Hence, in the following all $\chi^2$ are
computed assuming a diagonal covariance matrix; we refer the reader to
the appendix \ref{ApenB} for a discussion of the interpretation of
this {\em diagonal} $\chi^2$ and for more details on the procedure we
have followed to take into account the correlation among the data (in
time or in distance, see below) and the way we have computed the
statistical errors on the values of the critical
exponents.~\footnote{The same fit performed with the help of
  Gnuplot~\cite{Gnuplot} (with a diagonal covariance matrix) provides
  a $z= 2.012$ with an asymptotic error of $0.00075$.  In order to
  obtain the right statistical error, we need to divide this
  asymptotic error by $\sqrt{\chi^2/\mathrm{d.o.f.}}=0.687$~\cite{Young},
  obtaining the final value of $z=2.012(1)$. Notice that the computed
  error discarding correlations among the different times is a factor 13 times
  smaller.}

\begin{figure}[h!]
\centering
\includegraphics[width=\columnwidth, angle=0]{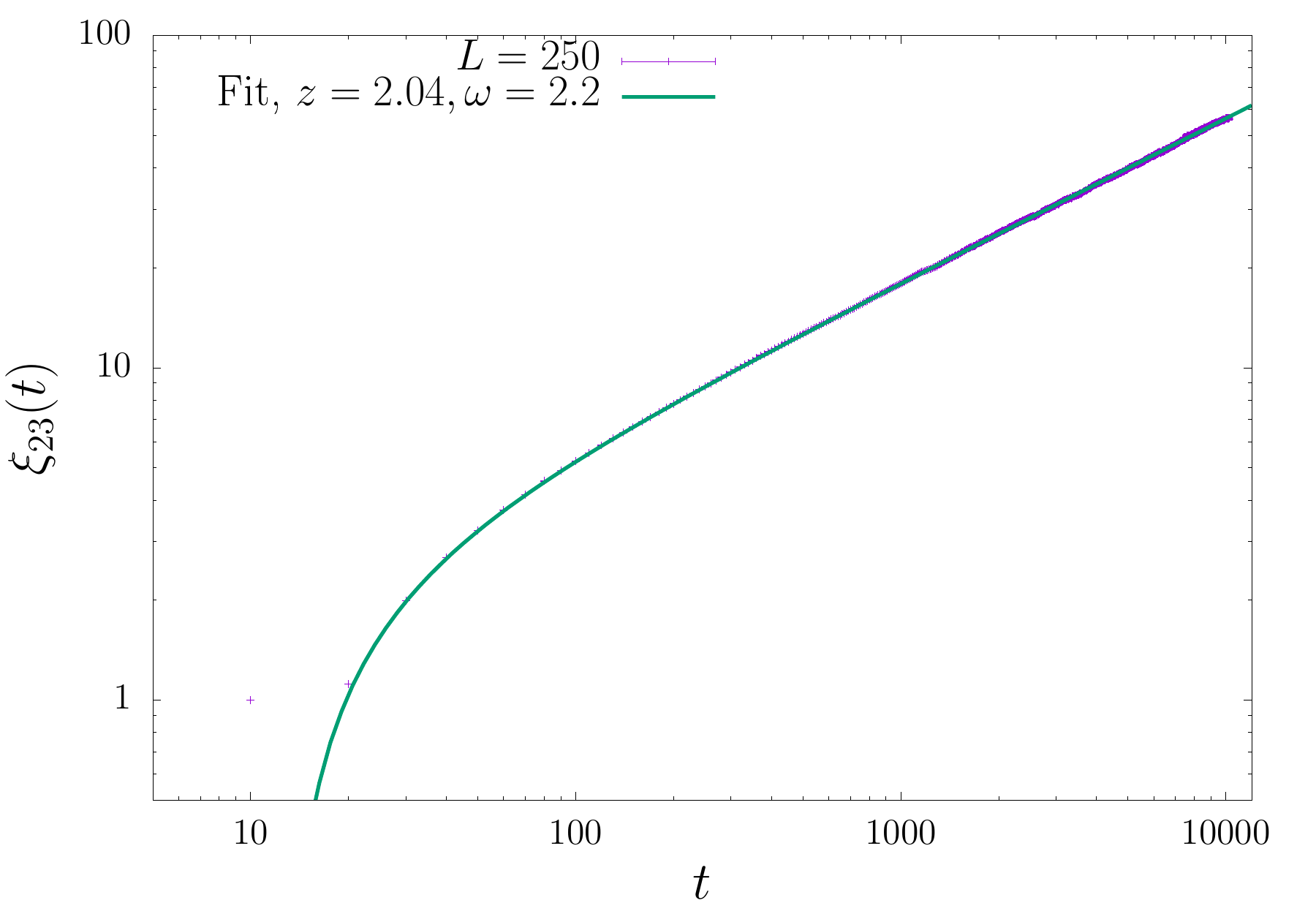}
\caption{(color online) Behavior of the dynamic correlation length for
  $L=250$ in the out of equilibrium regime. The fit is only for the $L=250$ data in
  the region $100\le t<8100$.}
\label{fig:XI12PURE}
\end{figure}

Having computed $\xi_{23}$ and $I_2 \propto \xi_{23}^{2-\eta}$ we can,
as a check, estimate the $\eta$ exponent. Fig.~\ref{fig:I2PURE} shows
$I_2$ as a function of $\xi_{23}$. We can compute $\eta$ using the
time interval $t\in[100,8100)$ obtaining $\eta=0.029(20)$ and
  $\omega=0.8(4)$ with $\chi^2/\mathrm{d.o.f.}=342/796$. In this
  case the $\omega$-exponent is similar to the equilibrium value
  $\omega \simeq 0.78$.~\cite{Guida:98,Hasen:01} We can improve the
  value of $\eta$ by fixing the correction-to-scaling exponent
  $\omega$ to the equilibrium value $\omega=0.78$, obtaining
  $\eta=0.044(7)$ with $\chi^2/\mathrm{d.o.f.}=666/757 $ ($t\in
  [50,8100)$).  Our value compares very well (but with 20 times more
    error) with that computed at equilibrium:
    $\eta=0.0378(3)$.~\cite{Campos:02,Hasen:11} Finally, neglecting
    the correction-to-scaling term, we have found $\eta=0.043(6)$ with
    $\chi^2/\mathrm{d.o.f.}=676/768$ ($t\in [400,8100)$). All three
      reported values are statistically compatible.

\begin{figure}[h!]
\centering
\includegraphics[width=1.0\columnwidth, angle=0]{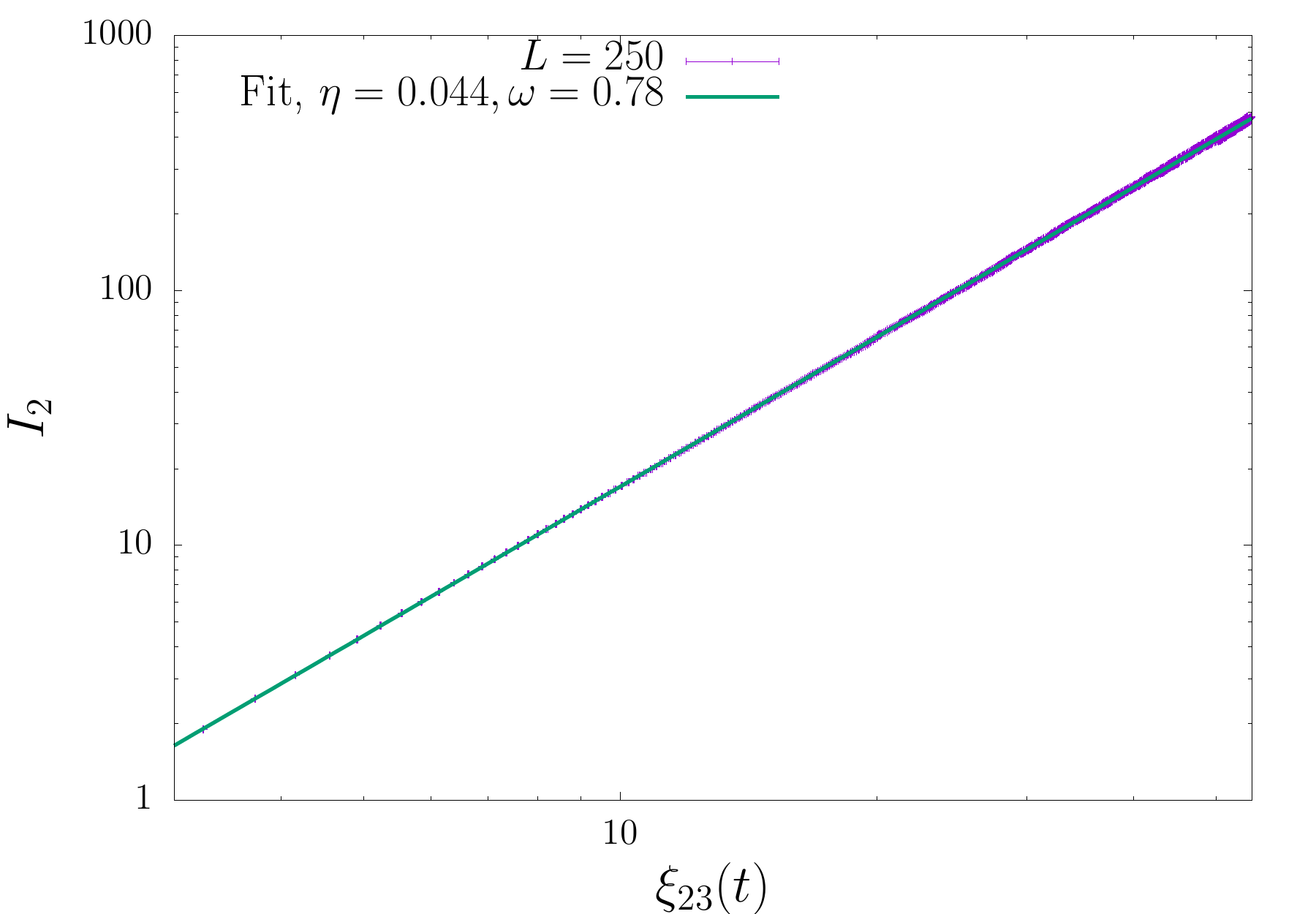}
\caption{(color online) Behavior of $I_2(\xi_{23})\propto \chi$ for
  $L=250$. We also plot the fit taking into account scaling
  corrections but with the $\omega$-exponent fixed to the
  equilibrium value as described in the text.}
\label{fig:I2PURE}
\end{figure}

\subsection{Correlation function for larger times}
\label{sec:eta}

In Fig. \ref{fig:CPURELARGECROSS} we plot $C(r,t)$ for different times
using $L=128$ data (200 initial conditions) and very long times. One
can see the crossover of the dynamic correlation function between the
off-equilibrium regime and the equilibrium one. In appendix
\ref{ApenB} we provide more details about the functional form of
$C(r,t)$ in the out of equilibrium regime.  We can also check that we
have reached the equilibrium regime by plotting the behavior of
$\xi_{23}(t)$ (see $L=128$ curve of Fig. \ref{fig:XI12EQUIL}). This
non-local observable has clearly reached its equilibrium (plateau)
value. We can safely assume that for $t>4\times 10^5$ we have
thermalized the $L=128$ lattice and we can try to extract the value of
the the anomalous dimension by averaging the correlation function
above this time.

The analytical behavior at the critical point in this regime (large
$L$) is given by Eq. (\ref{eq:corre-equi}).  Having in mind that we
are using periodic boundary conditions, we can write the following
improved equation to fit our numerical data
\begin{equation}
  C(r,L) =\frac{A}{r^{1+\eta}}+ \frac{A}{(L-r)^{1+\eta}}\,.
  \label{eq:fitcorre}
\end{equation}
By fitting the data of Fig. \ref{fig:CPURELARGE} to this functional
form, we obtain $\eta=0.026(4)$ (by using only $t>4\times 10^5$, $r\ge
16$ and $\chi^2/\mathrm{d.o.f}=44/48$) in a good statistical
agreement with the value drawn from equilibrium studies
$\eta=0.0378(3)$.  We have followed the method described in appendix
\ref{ApenB} in order to obtain the error in the $\eta$
exponent.~\footnote{The same fit, assuming no correlation among the
  different values of the correlation function, provides an error of
  0.0013, three times smaller than that obtained in our procedure.}

\begin{figure}[h!]
\centering
\includegraphics[width=\columnwidth, angle=0]{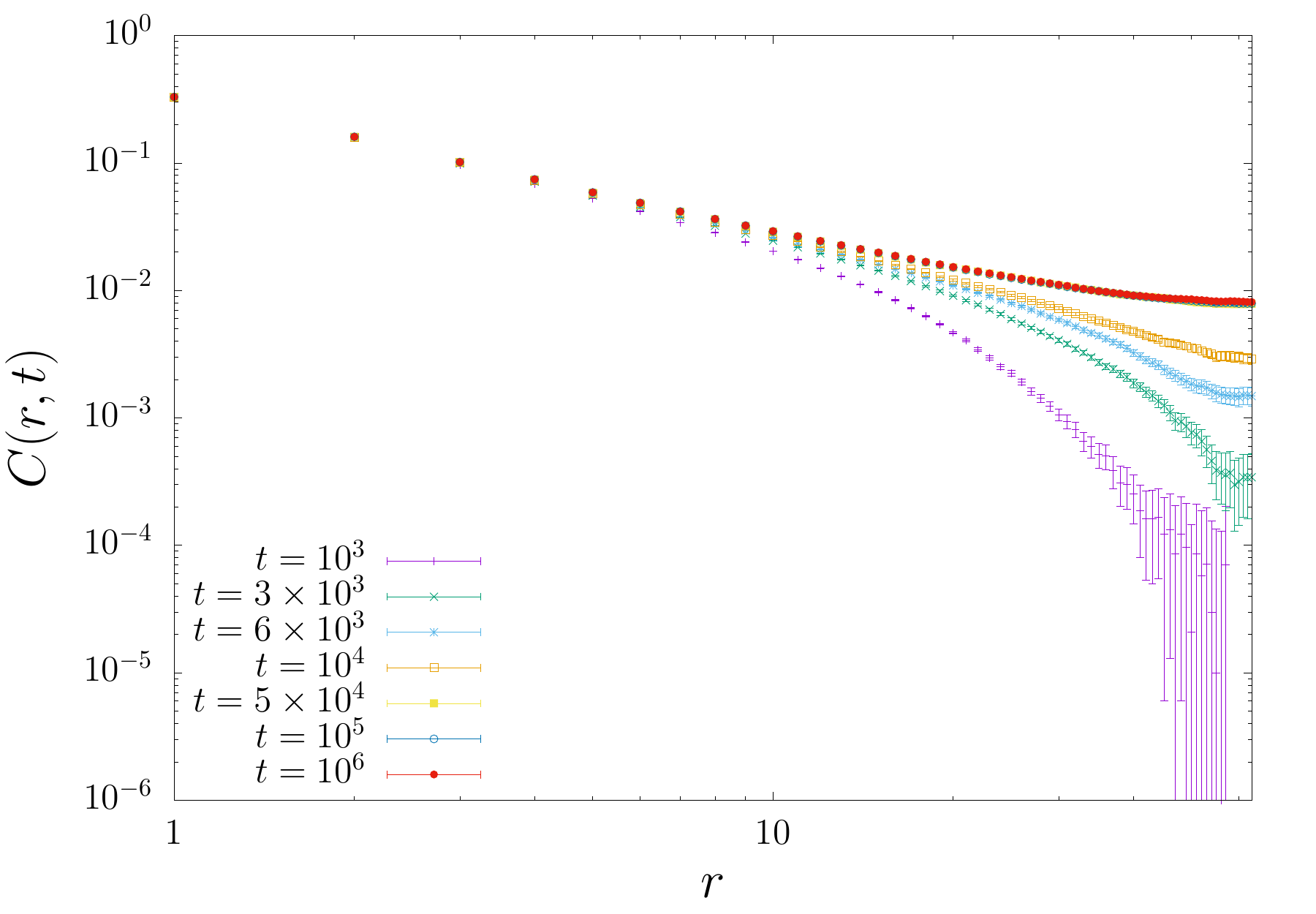}
\caption{(color online) Correlation function at criticality for a
  $L=128$ lattice. We have drawn different times in order to show the
  crossover between the out of equilibrium region and the equilibrium
  one. Notice the bad signal-noise ratio in the tail of $C(r,t)$ for
  large $r$ and shorter times $t$, and how this ratio improves with
  time, generating a plateau (due the periodic boundary conditions)
  with small error.}
\label{fig:CPURELARGECROSS}
\end{figure}

\begin{figure}[h!]
\centering
\includegraphics[width=\columnwidth, angle=0]{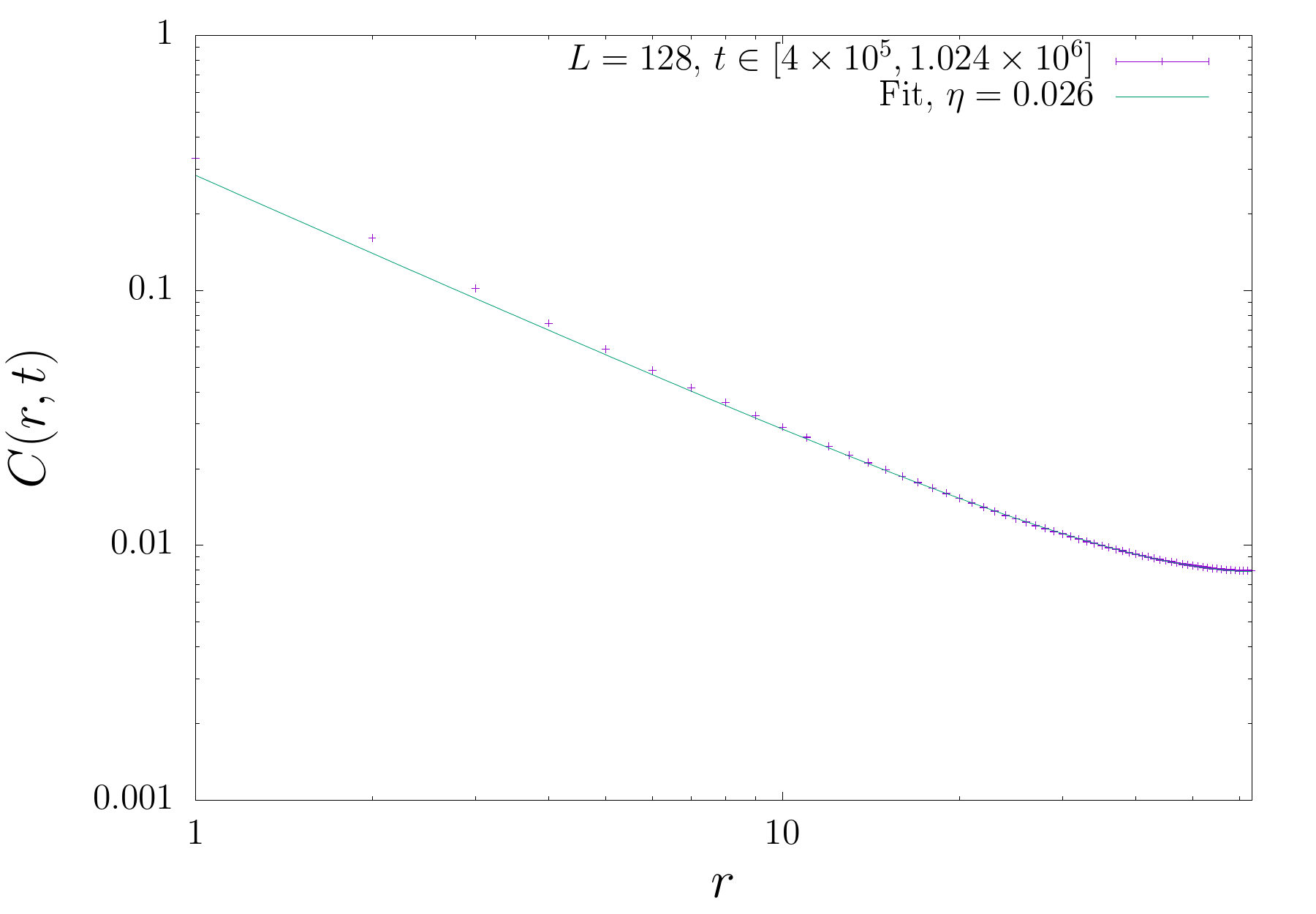}
\caption{(color online) Equilibrium correlation function at
  criticality for a $L=128$ lattice. The continuous line is a fit to
  Eq. (\ref{eq:fitcorre}) with $\eta=0.026$.}
\label{fig:CPURELARGE}
\end{figure}

\subsection{Energy for larger times}
\label{sec:ener}

\begin{figure}[h!]
\centering
\includegraphics[width=\columnwidth, angle=0]{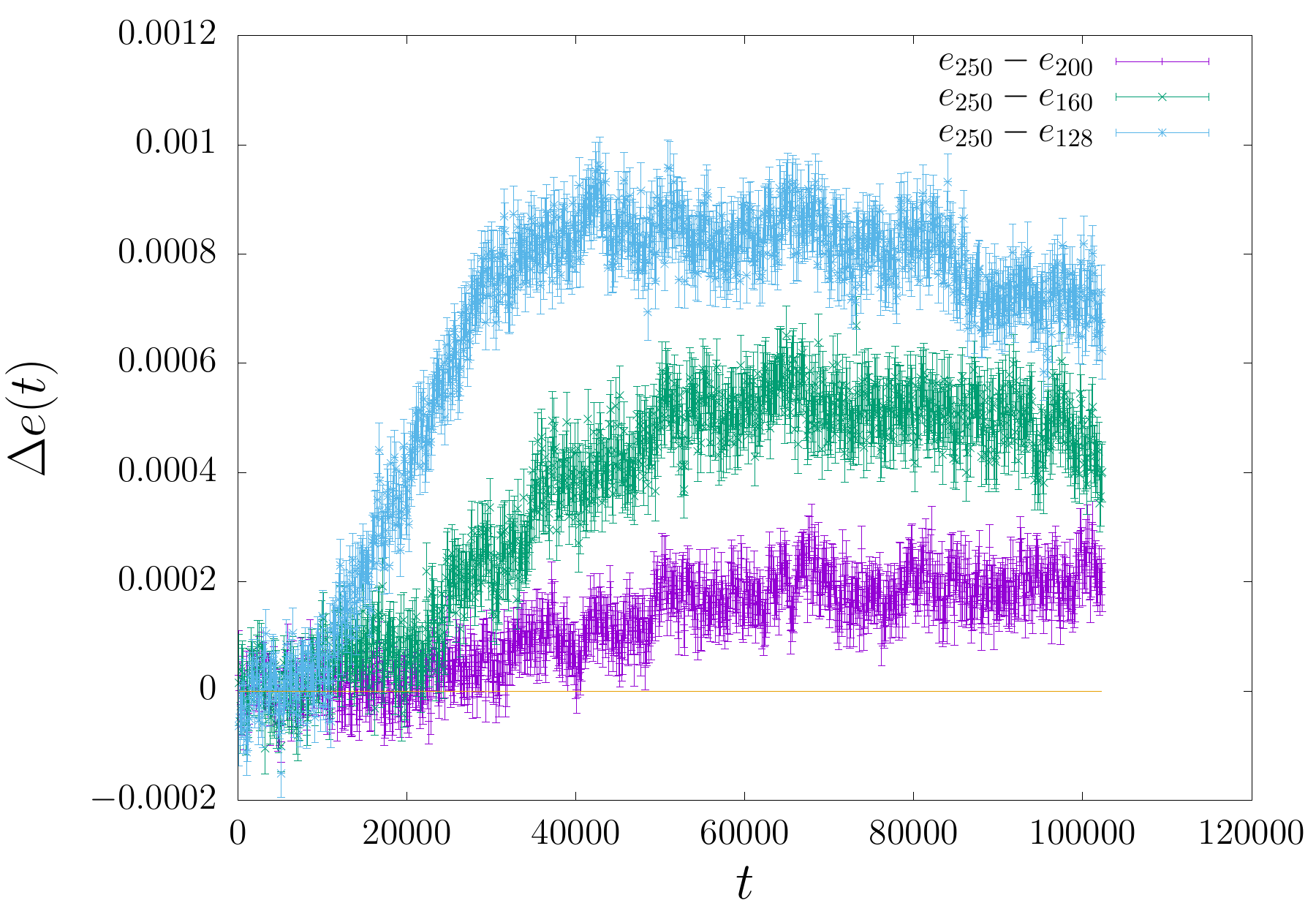}
\caption{(color online) We show the difference of energies, $\Delta
  e(t)$ for three pairs of lattice sizes as a function of time:
  $e_{L=250}-e_{L=200}$, $e_{L=250}-e_{L=160}$ and 
  $e_{L=250}-e_{L=128}$. The zero value has been marked with a
  horizontal line. Notice that the $L=250$ data are asymptotic (as
  compared with those of $L=200$) for $t<48000$ (the data are at one
  standard deviation of the zero value).}
\label{fig:enerL}
\end{figure}

We have analyzed the behavior of the energy at criticality in order to
compute the ratio of critical exponents $(d-1/\nu)/z$. To analyze this
behavior, we have run $L=128$ (153 initial conditions, i.c. in the
following), $L=160$ (600 i.c.), $L=200$ (684 i.c.) and $L=250$ (684
i.c.) for longer times $t<102400$.

Firstly, we study in Fig. \ref{fig:enerL} the effect of a finite size
lattice on the values of energy as a function of time. From this
figure one can see that it is safe to take fits only in the range
$t<48000$ in order to avoid finite size effects (at least in the
precision of our simulation).

In Fig. \ref{fig:ETCPURE} we show the results for the largest lattice
$L=250$.  We have fitted the $L=250$ data to a power law, in the time
interval $t\in [1000, 48000)$ obtaining $z=2.034(22)$ and
$e_\infty=-0.989505(17)$, with a diagonal
$\chi^2/\mathrm{d.o.f.}=985/939$. We have fixed in the fit the value
$\nu=0.7117(5)$.~\cite{Campos:02,Hasen:11} The really small error bar
of the $\nu$ exponent has not a measurable effect in the final error
bar of $z$. 
To finish the analysis of the energy, we have also checked
corrections to scaling for this observable and we have found that the
exponent $\omega_\mathrm{eff}=2\times 0.78$ describes very well the
numerical data obtaining $z=2.13(7)$ and $e_\infty=-0.989525(22)$ with
$\chi^2/\mathrm{d.o.f}=980/947$.

\begin{figure}[h!]
\centering \includegraphics[width=\columnwidth,
  angle=0]{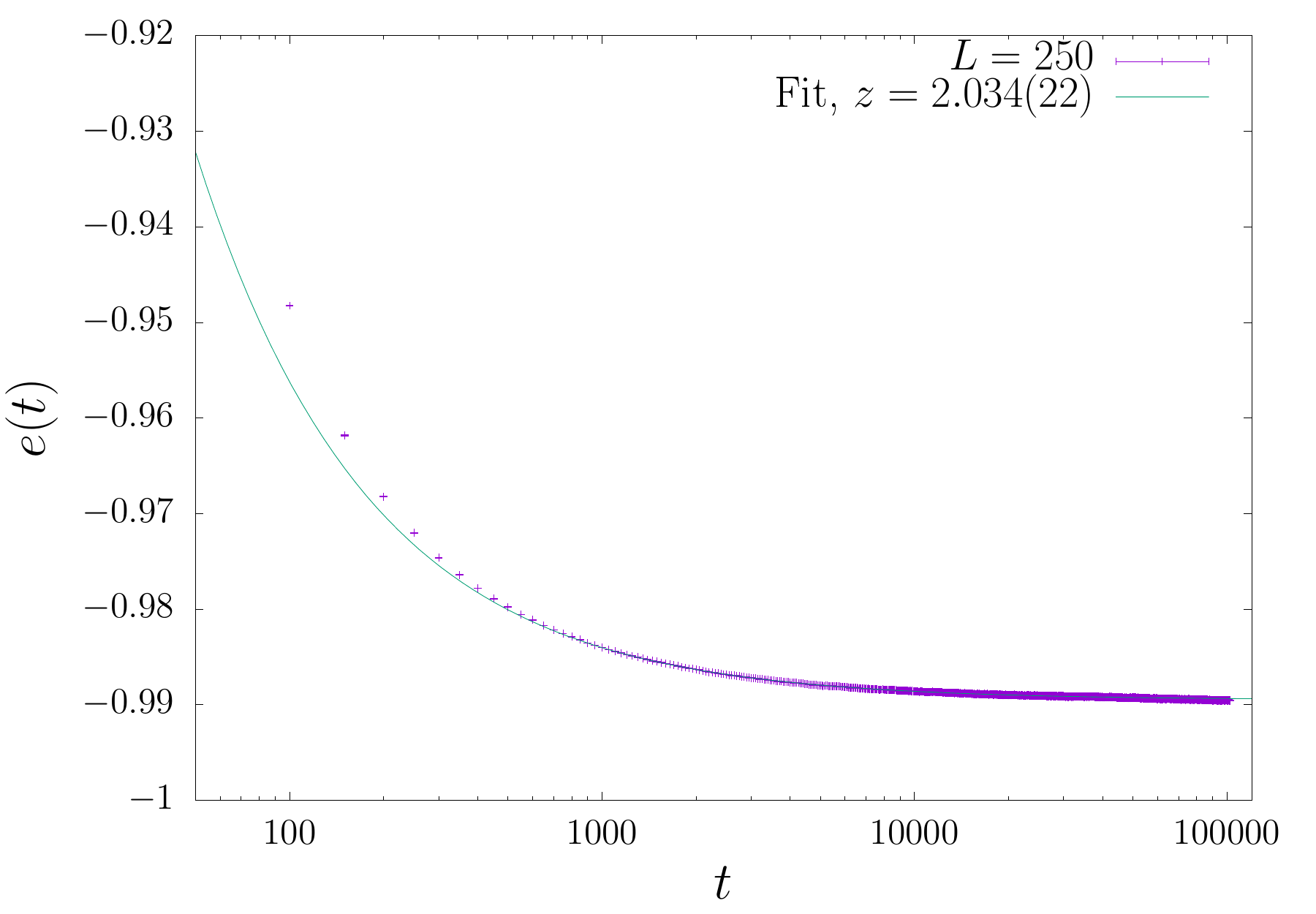}
\caption{(color online) Behavior of the energy $e(t)$ at the critical
  point for the $L=250$ run. We draw also the fit in order to extract
  the ratio $(d-1/\nu)/z$ with $d=3$, $\nu=0.7117(5)$ fixed getting
  $z=2.034(22)$.}
\label{fig:ETCPURE}
\end{figure}
  
\section{Conclusions}
\label{sec:concl}

By performing in and out equilibrium numerical simulations we have
computed the dynamic critical exponent $z$.

The most accurate value has been computed in the equilibrium regime by
studying the integrated correlation time as a function of the lattice
size: $z=2.033(5)$. We have found a correction-to-scaling exponent
$\omega=2.7(3)$. In addition we have provided strong numerical
evidences about the proportionality of the integrated and exponential
correlation times.

We have also computed the $z$ exponent in the out-of-equilibrium regime
obtaining $z=2.04(2)$ and $\omega=2.2(4)$ which is similar to the
$\omega$-exponent computed at equilibrium.  Moreover, we have checked
the consistency of the computed critical exponents at equilibrium with
the out of equilibrium ones with and without considering corrections
to scaling. The (equilibrium) value of $\nu$ provides us, by
monitoring the energy, with a another dynamic exponent estimate
($z=2.034(22)$) fully compatible with the previous ones.

Furthermore, our value of $z$ has improved the statistical precision
of that computed in numerical simulations performed at equilibrium in
relatively small lattices ($L\le 24$).~\cite{Peczak:93} Our computed
values match very well with that obtained in experiments and with the
exponent computed using field theoretical
techniques~\cite{Antonov:84} (although in this framework it is very
difficult to assign an uncertainty to this estimate).

\acknowledgments

We thank L. A. Fernandez, M. Lulli, A. Pelissetto, V. Martin-Mayor,
J. Salas, J. A. del Toro and D. Yllanes for discussions.  This work
was partially supported by Ministerio de Econom\'{\i}a y
Competitividad (Spain) through Grant No.\ FIS2016-76359-P, by Junta de
Extremadura (Spain) through Grant No.\ GRU10158 and IB16013 (partially
funded by FEDER). We have run the simulations in the computing
facilities of the Instituto de Computaci\'{o}n Cient\'{\i}fica
Avanzada (ICCAEx) and in the CETA-Ciemat thanking Dr. A. Paz for his
support.

\appendix

\section{Details of the numerical simulations and GPU parallel implementation}
\label{ApenA}

We have simulated the Heisenberg model using the Metropolis Algorithm
on CPUs and GPUs (see Table \ref{hardware_features}). We have
simulated $L=128$, 160, 200 and 250 for more than 10000 random initial
conditions.  The GPU code has been programmed in CUDA
C.~\cite{nvidia11a} The original C code which simulates the Heisenberg
model has been parallelized in three parts:

\begin{enumerate}

\item Computation of the nearest neighbors of each spin: the C
code has a loop which goes sequentially through all the spins 
one by one. 
However, in the GPU code each spin has associated an 
execution thread and all the nearest neighbors 
of every spin are computed at once.

\item Metropolis Algorithm: in the sequential C code we can
find several loops in the Metropolis part. So, the parallel GPU
code reduces meaningfully the execution time especially in large
systems ($L \sim 200$).
Moreover, the lattice 
has been divided using a checkboard scheme 
(Fig. \ref{checkboard_scheme}).~\cite{Lulli:15a} In this way, 
the Metropolis algorithm has been executed first of all in the 
``white'' spins and after that in the ``black'' ones.

\item Random numbers: to have high quality random numbers is 
mandatory in Computational Physics. Initially, we have used the
CURAND random numbers which are part of the CUDA C distribution.
~\cite{nvidia11a} The problems with the CURAND random numbers 
have appeared when we have performed long simulations using a
huge quantity of random numbers. To avoid these problems we have
used Congruential Random Numbers.~\cite{Knuth:98a}

\end{enumerate}

\begin{table}
\begin{ruledtabular}
\begin{tabular}{cccc}
CPU/GPU & CPU Intel & GPU Geforce  & GPU Tesla \\
model   & Core i7   & GTX 1080 G1  & K80 \\   

\hline

Cores & 20 & 2560 & 4992 \\
Core clock & 2.26 GHz & 1.86 GHz & 0.88 GHz \\
Total memory & 24 GB & 8 GB & 24 GB \\  
Memory & - & - & 480 GB/s \\
bandwidth & & & \\

\end{tabular}
\end{ruledtabular}
\caption{Hardware features of the CPUs and GPUs.\label{hardware_features}}
\end{table}  

\begin{figure}[h!]
\centering
\scalebox{1}{\includegraphics{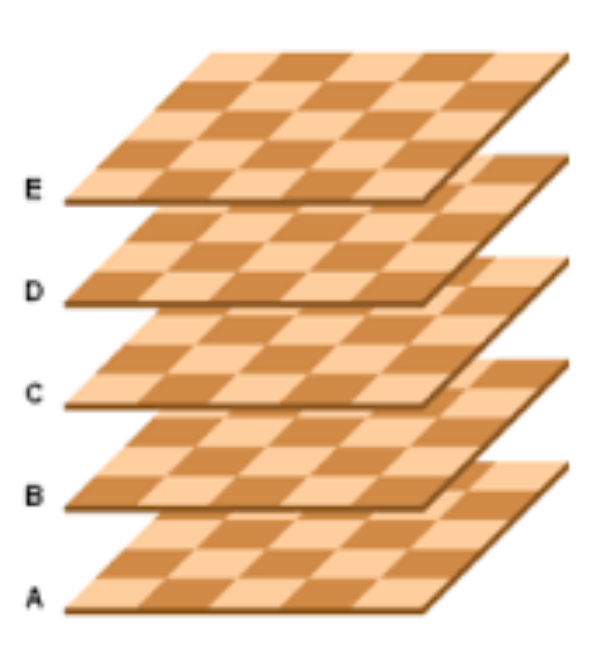}}
\caption{(color online) Division of the three dimensional lattice
using a checkboard scheme.}
\label{checkboard_scheme}
\end{figure}

Making use of the GPU Tesla K80 we have achieved a speedup of 22 
which represents an important reduction of the execution time.

\section{Details of the analysis of the computation of the correlation length}
\label{ApenB}

We describe the different steps we have followed in order to
compute $\xi(t)$ and its associated exponent
$z$.~\cite{Janus1,Janus2,ThesisDavid,Michael:94} The important point
of this approach is to avoid the use of the full covariance matrix
since this matrix is frequently singular (see for
example~\cite{ThesisDavid,Seibert:94}). Thus, our  procedure is the
following:

\begin{enumerate}

  \item We compute using the jackknife method over the set of the initial
    conditions, the statistical  error of  $C(r,t)$, denoted
    as $\sigma[C(\Lambda,t)]$.

  \item To compute $I_k$ we introduce a cutoff to
    have a good control of the signal to noise ratio of $C(r,t)$ for
    large values of $r$ (see also Fig.~\ref{fig:CPURELARGECROSS}).

    \begin{itemize}

    \item We compute the cutoff $\Lambda$ using the condition $\sigma[C(\Lambda,t)]=
      4 C(\Lambda,t)$.

    \item For a fixed $t$ and $r_\mathrm{min}< r < \Lambda$ we fit the
      correlation function to the functional form given by
      \begin{equation}
        C(r,t)=\frac{a_1}{r^{a_2}} \exp(-a_{3} r^{a_4})\,.
        \label{eq:corre}
      \end{equation}
      with $r_\mathrm{min}$ is the minimum value of $r$ which
      provided, for $C(r,t)$, a good fit (e.g. $\chi^2/\mathrm{d.o.f.}
      \sim 1$) to Eq.(\ref{eq:corre}). In Fig.  \ref{fig:corre} we
      report the dependence of the exponents $a_2$ and $a_4$ with the
      Monte Carlo time. Notice that $a_2$ converges to the equilibrium
      value (see Eq. (\ref{eq:corre-equi})) given by
      $1+\eta=1.0378$ and $a_4\simeq 1.8$.

      \begin{figure}[h!]
        \centering
        \includegraphics[width=\columnwidth, angle=0]{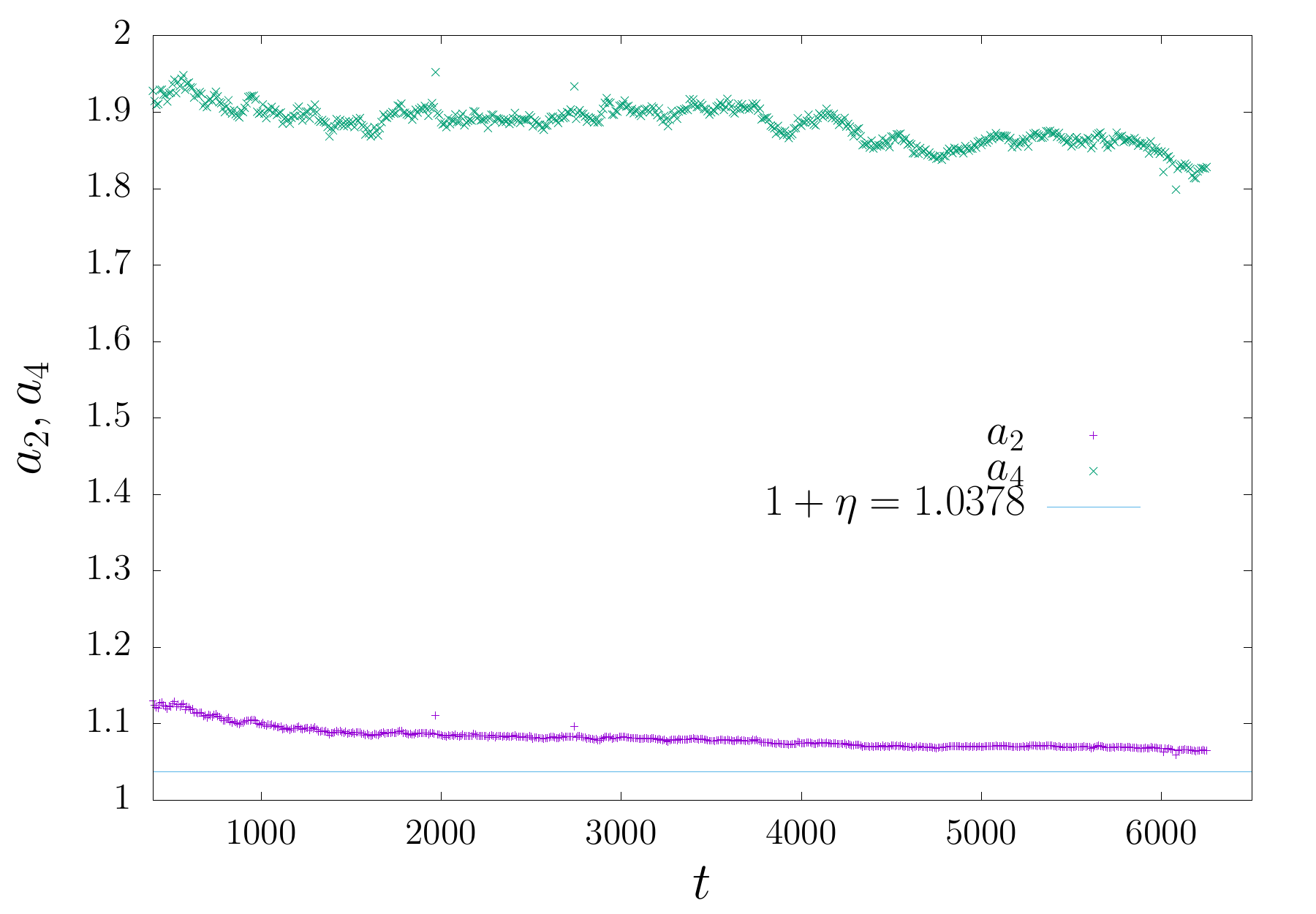}
        \caption{(color online) Behavior of the exponents $a_2$ and $a_4$ as a
          function of time for $L=200$. The horizontal line is the equilibrium theoretical
          expectation for $a_2$, namely $1+\eta=1.0378$.}
        \label{fig:corre}
      \end{figure}

    \item  We compute the integral in
      Eq. (\ref{eq:I}) using the numerical values of the correlation
        $C(r,t)$ for $r<\Lambda$ and using the values provided by the
      fit (Eq. (\ref{eq:corre})) for $\Lambda < r <L/2$.
      
    \item Using the previous procedure, we compute the statistical
      error of $\xi (t)$ using again the jackknife method over the set of the
      initial conditions. The time interval for the fit is decided by
      imposing a diagonal $\chi^2/\mathrm{d.o.f}\sim
      1$.

    \item The jackknifed $\xi$'s are used to compute the jackknifed
      values of $z$ and this allows us to compute the statistical
      error of the dynamic critical exponent using the standard
      deviation in the jackknife method. Notice that for extracting
      $z$ on each jackknife block, we use the {\em diagonal} covariance
      matrix. However, the jackknife procedure reproduces with high
      accuracy the effect of the correlations among the different
      times. 
      
    \end{itemize}
    
    Notice that the {\em diagonal} $\chi^2/\mathrm{d.o.f.}$ has not a
    rigorous interpretation as that of the full (non diagonal) one. One can
    show (see the detailed analysis of this procedure carried out in
    section B.3.3.1 of Ref. ~\cite{ThesisDavid}) that the
    {\em diagonal} $\chi^2/\mathrm{d.o.f.}$ behaves as if there were a
    small number of degrees of freedom, hence, one can not compute
    confident limits as usual.

    Finally, in Ref.~\cite{Lulli:16} was
    shown that the error bars are essentially equal (using this
    jackknife procedure, neglecting the correlations among the
    data) to those obtained taking into account all the statistical
    correlations among the data.
     
\end{enumerate}

\end{document}